\documentclass[useAMS,usenatbib]{mn2e}
\usepackage{natbib}
\usepackage{subfigure}
\usepackage{gensymb}
\usepackage{amssymb}
\usepackage{wasysym}
\usepackage{graphicx}
\usepackage{color}

\newcommand{\z}{\color{black}}

\DeclareMathSymbol{\varOmega}{\mathord}{letters}{"0A}
\DeclareMathSymbol{\varSigma}{\mathord}{letters}{"06}

\DeclareMathSymbol{\varPsi}{\mathord}{letters}{"09}

\newcommand{\lsim}{\mathrel{\rlap{\lower4pt\hbox{\hskip1pt$\sim$}}
   \raise1pt\hbox{$<$}}}                
\newcommand{\gsim}{\mathrel{\rlap{\lower4pt\hbox{\hskip1pt$\sim$}}
   \raise1pt\hbox{$>$}}}                

\begin{document}
\title{Eight billion asteroids in the Oort cloud}
\author[Shannon et al.]{Andrew Shannon$^{1}$, Alan P. Jackson$^{2}$, Dimitri Veras$^{3}$, \& Mark Wyatt$^{1}$ \\
$^{1}$Institute of Astronomy, University of Cambridge, Madingley Road, Cambridge, CB3 0HA, UK \\
$^{2}$School of Earth \& Space Exploration, Arizona State University, 781 E Terrace Mall, Tempe, AZ, 85287-6004, USA \\
$^{3}$Department of Physics, University of Warwick, Gibbet Hill Road, Coventry, CV4 7AL, UK}

\maketitle

\begin{abstract}
The Oort cloud is usually thought of as a collection of icy comets inhabiting the outer reaches of the Solar system, but this picture is {\z incomplete}.  We use simulations of the formation of the Oort cloud to show that $\sim 4\%$~of the small bodies in the Oort cloud should have formed within 2.5 au of the Sun, and hence be ice-free rock-iron bodies.  If we assume these Oort cloud asteroids have the same size distribution as their cometary counterparts, the Large Synoptic Survey Telescope should find roughly a dozen Oort cloud asteroids during ten years of operations.  Measurement of the asteroid fraction within the Oort cloud can serve as an excellent test of the Solar system's formation and dynamical history.  Oort cloud asteroids could be of particular concern as impact hazards as their high mass density, high impact velocity, and low visibility make them both hard to detect and hard to divert or destroy.  However, they should be a rare class of object, and we estimate globally catastrophic collisions should only occur about once per billion years.  
\end{abstract}


\section{Introduction}
The Sun is surrounded by a cloud of hundreds of billions of comets, with semimajor axes of $10^4 - 10^5~\rm{au}$.  They did not form at such distances, but rather within a few tens of au of the Sun, and were scattered onto large $\left(10^4~\rm{au} \lesssim a \lesssim 10^5~\rm{au}\right)$, highly eccentric $\left(e \sim 1\right)$~orbits~by the planets \citep{1950BAN....11...91O,2008ssbn.book..315D}.  At distances of thousands of au or more from the Sun, perturbations from the Galactic tide and nearby stars raised the perihelia of these comets to beyond the reach of Neptune, allowing their orbits to be stable for billions of years \citep{1986Icar...65...13H,1987AJ.....94.1330D,2004ASPC..323..371D}.

As the Solar system is a complex dynamical system of eight planets and numerous small bodies and satellites, and as it is perturbed by passing stars \citep{1950BAN....11...91O}, giant molecular clouds \citep{1978bs...symp..327B}, and the Galactic tide \citep{1986Icar...65...13H}, all of which may be important for the evolution of Oort cloud comet orbits, the problem of the formation of the Oort cloud is typically approached with N-body simulations.  It is generally believed that the planets have migrated during the Solar system's history \citep[e.g.,][]{1984Icar...58..109F,1993Natur.365..819M,2005Natur.435..459T,2011Natur.475..206W}, although the details of these scenarios are uncertain.  Likewise, the birth environment of the Sun \citep{2010ARA&A..48...47A} and its subsequent migration through the galaxy \citep{2002MNRAS.336..785S,2008ApJ...684L..79R} are unknown.  The properties of the Oort cloud can be used to constrain the dynamical history of the solar system's planets \citep[e.g.,][]{2013AJ....146...16L,2013Icar..225...40B,2014Icar..231..110F} and the solar neighbourhood over its history \citep[e.g.,][]{2010Sci...329..187L,2011Icar..215..491K,2012Icar..217....1B}, but {\z different combinations of the models of} the dynamical history of the solar system, the history of external perturbers, and the initial populations of small bodies {\z can produce the same total population of Oort cloud comets.}

To break this degeneracy, one approach is to try to associate Oort cloud objects with their birth region.  {\z A similar approach has already been suggested for Main belt asteroids \citep{2014Natur.505..629D}.}
Traditionally, small bodies have been divided between comets (icy bodies on highly eccentric orbits), and asteroids (rock-iron bodies on lower eccentricity orbits).  
{\z However, while these properties were well correlated observationally, the outgassing of cometary nuclei increases their brightness by orders of magnitude, making them much easier to discover; meanwhile, objects in the asteroid belt had gigayears to outgas their surface volatiles, more than enough time to reduce or eliminate any activity \citep{1980A&A....85..191W,1997Icar..127...13L,2008ApJ...682..697S}.} 
Questioning this assumption is motivated by the discovery that the separation between asteroids and comets is not a clean one, with icy, outgassing bodies found in the main asteroid belt (called Active Asteroids or Main-belt Comets) \citep{2006Sci...312..561H, 2009M&PS...44.1863H,2013ApJ...771L..36J}, and the overlap between the orbital distributions of comets and asteroids \citep{2014Icar..234...66T}.  Similarly, modelling the dynamics of small bodies from the inner reaches of the Solar system has found that they are commonly ejected \citep{2012MNRAS.425..657J}, so we should also expect near-ejections to produce rocky, ice-free bodies in the Oort cloud, which we will refer to as Oort cloud asteroids.  This possibility was considered by \citet{1997ApJ...488L.133W} to explain the appearance of 1996 PW, an object in a long-period comet orbit which did not display outgassing.  They inferred that $\sim 1\%$~of the Oort cloud should be from the region of 3.3 to 5.2 au, based on the simulations of \citet{1987AJ.....94.1330D}.  Those simulations assumed that bound objects with aphelion $Q > 10^4~\rm{au}$~became comets, rather than modelling the effects of the Galactic tide and passing stars. Scaling {\z the 3.3-5.2 au result down to 1 au, \citet{1997ApJ...488L.133W}} estimated $2.3\%$~of the Oort cloud to have come from the region interior to Jupiter's orbit.  Follow-up observations found that although 1996 PW's appearance was compatible with a S- or D-type asteroid, it was also compatible with an extinct comet \citep{1998Icar..132..418D,2000Icar..143..354H}, and the issue of rocky bodies in the Oort cloud appears to have been subsequently neglected.

Modelling of the evolution of protoplanetary disks suggests the snow line may have been as close to the Sun as 0.7 au and as far as 10 au \citep{2007ApJ...654..606G,2011Icar..212..416M,2012MNRAS.425L...6M}~during its evolution.  However, observations of the population of active asteroids (sometimes called main belt comets) suggest that the snow line separation objects that formed with water ice and those that formed without it in the solar system is at $2.5 \sim 3.0~\rm{au}$ \citep{2011P&SS...59..365B,2011Icar..215..534S}.  Given this, Oort cloud asteroids from the region interior to 2.5 au could represent a significant fraction of bodies in the present day Oort cloud.

To address the question of how common asteroids should be in the Oort cloud, we perform simulations of the Oort cloud's formation, considering small bodies starting with dynamically cold orbits with semimajor axes from 0.5 au to 50 au.  In \textsection \ref{sec:model} we describe our numerical setup.  In \textsection \ref{sec:results}, we present the results of those simulations.  We estimate the detectability of such bodies in \textsection \ref{sec:detect}, their impact rate on the Earth in \textsection \ref{sec:earthimpacts}, and finish up with some discussion in \textsection \ref{sec:discussion}.

\section{Model}
\label{sec:model}

We perform simulations with the Mercury suite of N-body integrators \citep{1999MNRAS.304..793C}, augmented with the Galactic tidal model of \citet{2013MNRAS.430..403V} and the stellar flyby model of \citet{VSG}.  Interior to 6 au, simulations are performed with the outer seven planets (i.e., excluding Mercury), beginning with their present day\footnote{Midnight of July 6th, 1998} positions and velocities, with the ecliptic rotated to be inclined 67$\degree$~to the Galactic disk.  Integrations are performed with an eight day timestep.  For test particles at 6 au and beyond, we include only the outermost four planets, which allows us to use a 120 day timestep and somewhat lessen the computational burden.  Simulations are run for 4.5 Gyrs, or until all test particles have been removed.  Particles are removed by collisions with a planet or the Sun, or when they are more than 250 000 au from the Sun, exceeding the maximum extent of the Hill ellipsoid around the Sun \citep{2013MNRAS.430..403V}.

We use at least 1000 particles per au, except between 2.0 and 5.5 au, where the long-term stability of the asteroid belt and Jupiter trojans make this choice unfeasible, and beyond 30 au, where long-term stability again makes this unfeasible.  There we use at least 100 particles per au.  Bin widths are 0.1 au from 0.5 to 6.0 au, and 1.0 au thereafter.  Initial eccentricities are chosen randomly and uniformly from 0.00 to 0.05, and inclinations are chosen randomly and uniformly from 0 to $0.05 \frac{\pi}{2}$, and the other orbital angles randomly and uniformly from $0$~to $2\pi$.

\section{Results} 
\label{sec:results}

\begin{figure*}
 \centering
  \subfigure{\includegraphics[width=0.48\textwidth]{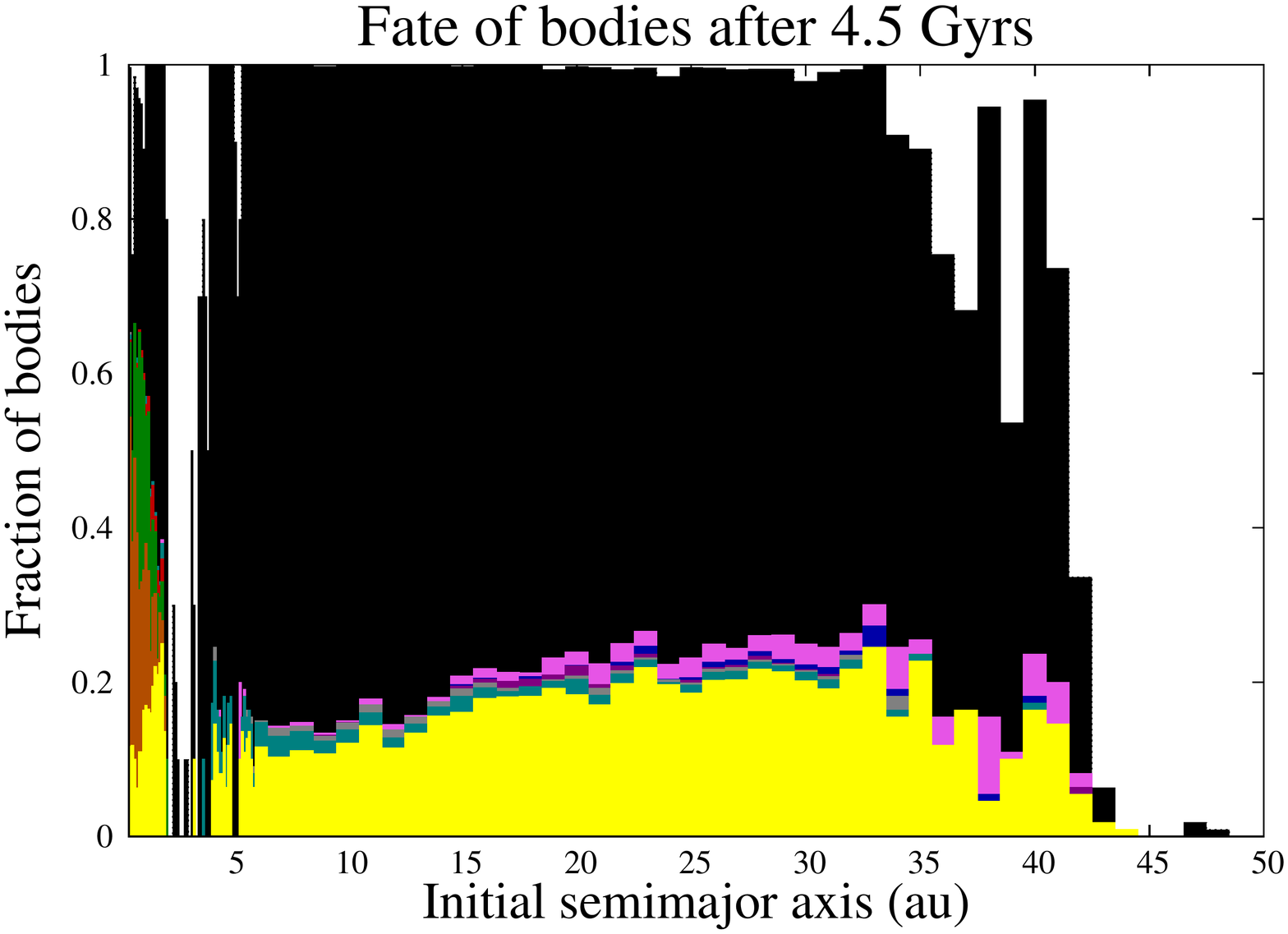}}
  \subfigure{\includegraphics[width=0.48\textwidth]{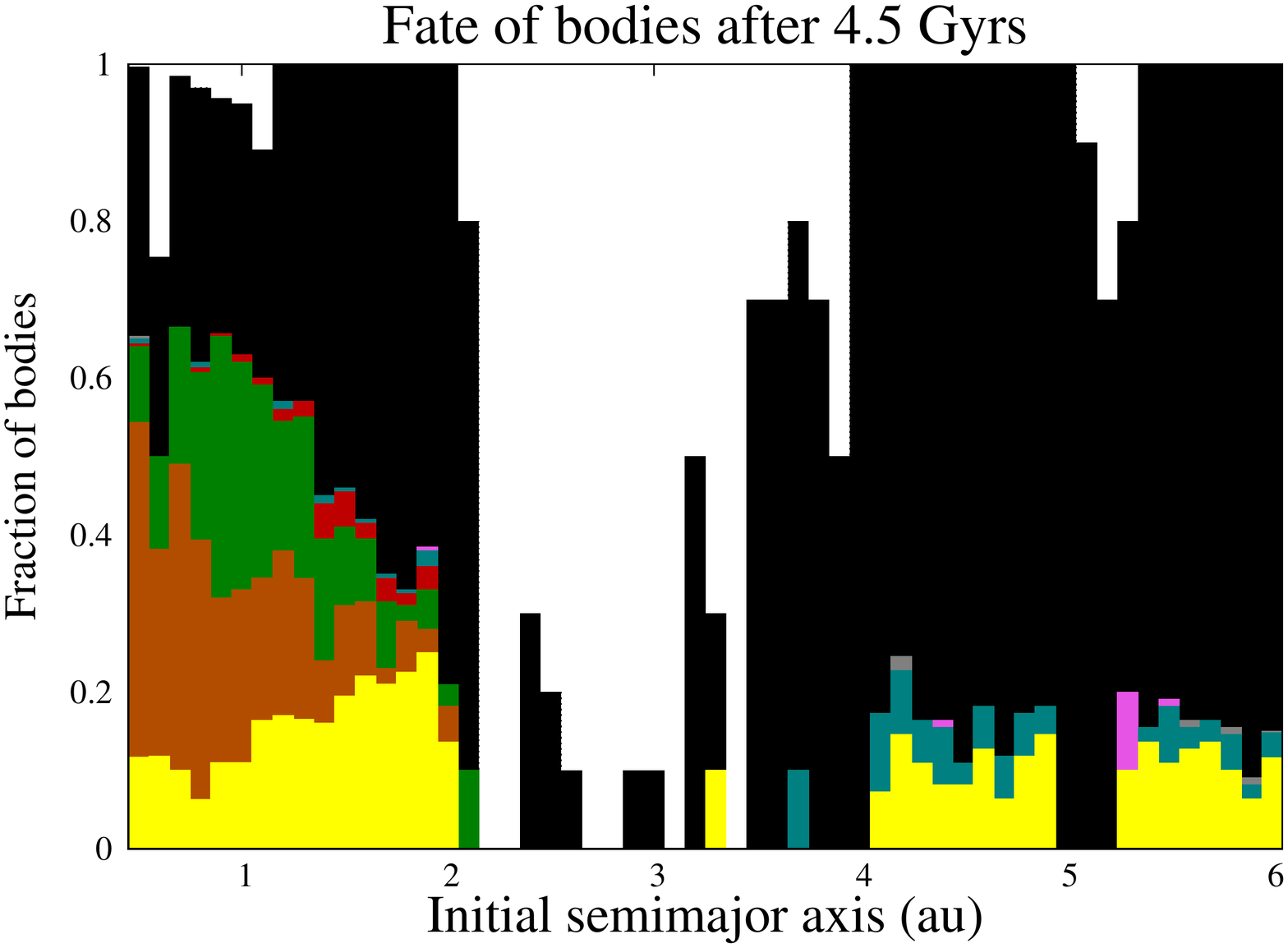}}
  \caption{The status of the test particles at $4.5$~Gyrs.  The left panel shows the entire solar system, while the right panels zooms in on the inner 6 au.  From bottom to top, the outcomes are impacted the Sun (yellow), impacted Venus (brown), impacted Earth (green), impacted Mars (red), impacted Jupiter (turquoise), impacted Saturn (grey), impacted Uranus (purple), impacted Neptune (blue), Oort cloud comet (pink), ejected (black), and otherwise still active (white).}
\label{fig:outcomes}
\end{figure*}

We plot the fate of particles after $4.5~\rm{Gyrs}$~of evolution in figure \ref{fig:outcomes}.  Beyond 1.5 au, ejection is the most common outcome for particles which are lost. Interior to $1.5~\rm{au}$, no single outcome dominates, with ejection, and collisions with Earth, Venus, and the Sun all important.  The majority of particles between $2.2~\rm{au}$~and $3.5~\rm{au}$~(i.e., the asteroid belt), and beyond $43~\rm{au}$~(i.e., the Main Kuiper belt)~are retained.  Significant numbers of Jupiter trojans, Neptune trojans, and inner Kuiper belt objects {\z \citep{1993AJ....105.1987H,1995AJ....110.3073D}}~are also retained.  We separate particles which are Oort cloud comets at $4.5 \times 10^9~\rm{years}$~from other surviving particles; we define particles to be members of the Oort cloud if they lie on an orbit with perihelion $q \geq 40~\rm{au}$~and semimajor axis $a \geq 1000~\rm{au}$; the former criterion selects those objects that are not interacting with the planets by close encounters; the latter criterion excludes Kuiper belt-type objects that begin with $q \geq 40~\rm{au}$~but have aphelia too small to interact with the Galactic tide or passing stars.

\begin{figure*}
 \centering
  \subfigure{\includegraphics[width=0.48\textwidth]{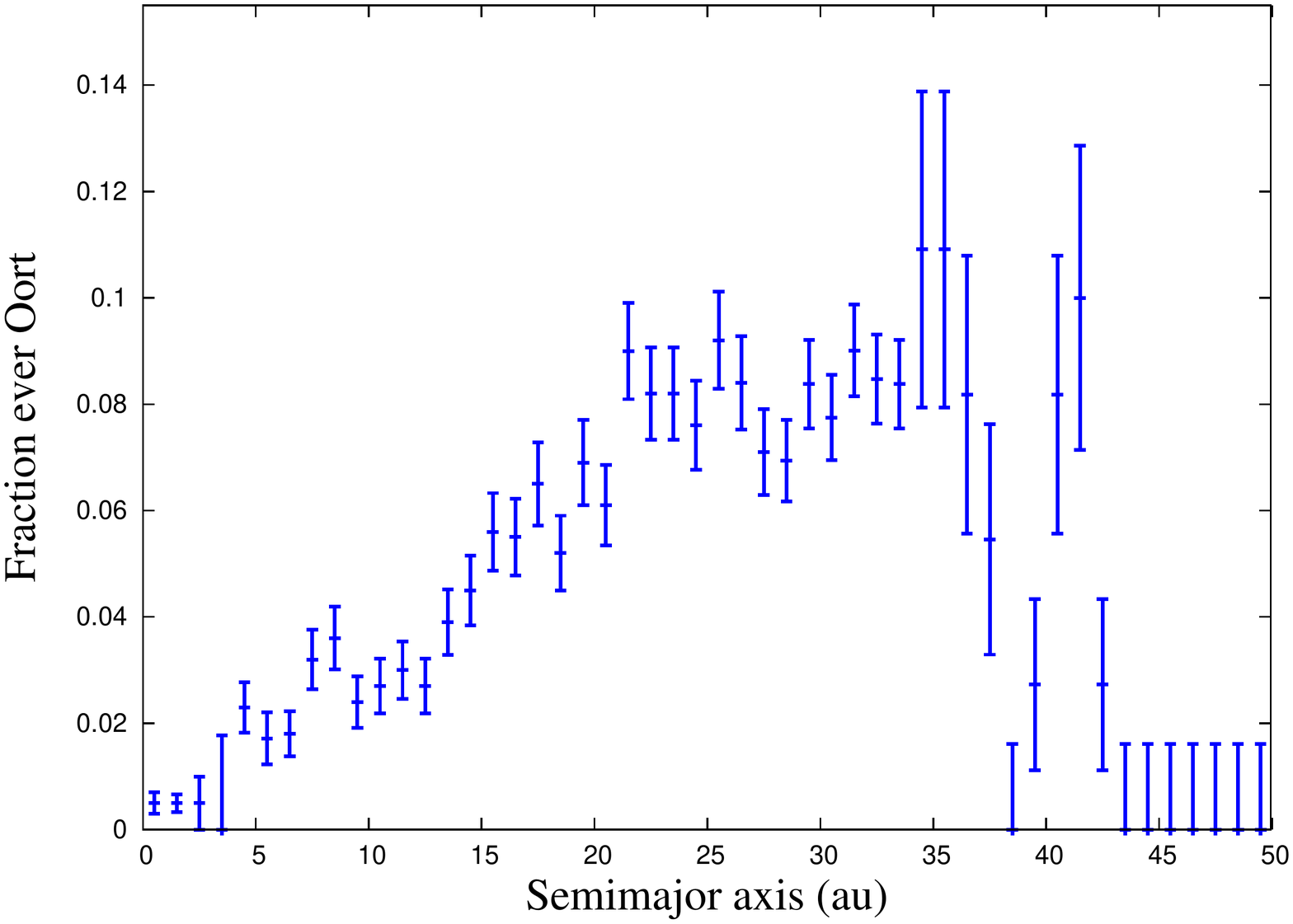}}
  \subfigure{\includegraphics[width=0.48\textwidth]{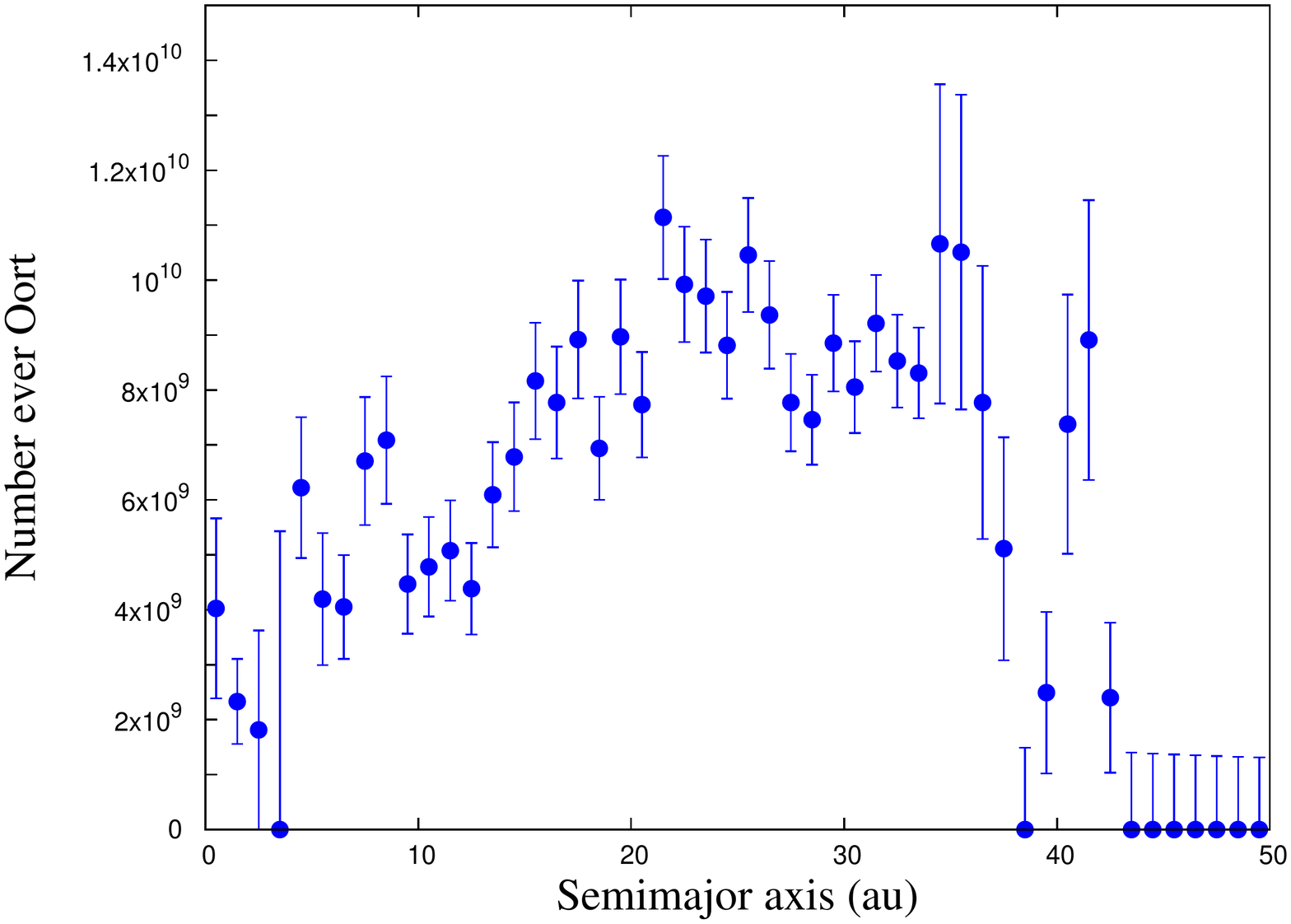}}
  \caption{In the left panel, we plot the fraction of all objects that were ever Oort cloud objects before $4.5$~Gyrs.  Only $\sim 25\%$~remain Oort cloud objects after 4.5 Gyrs of evolution.  In the right panel, we plot the total number of objects put into the Oort cloud, by assuming that the surface density of the objects begins with the same scaling as the Minimum Mass Solar Nebula, normalised so the total number of particles is $2 \times 10^{11}$~to be congruent with observations.}
\label{fig:oortrate}
\end{figure*}

Only about $25\%$ of our Oort cloud objects are retained until 4.5 Gyrs\footnote{Similar results are found by other groups \citep[e.g.,][]{2004SoSyR..38..325M,2008MNRAS.391.1350L,2011Icar..214..334F}}, so to obtain a bigger sample we focus on objects that are ever Oort cloud objects.  The fraction of objects that are ever Oort cloud objects is plotted in figure \ref{fig:oortrate} (left panel).  Interior to Jupiter's orbit, slightly less than $1 \%$~of objects ever become Oort cloud objects, rising to $\sim 10\%$~for objects beginning beyond the orbit of Uranus.  Qualitatively, this outcome is unsurprising, since \citet{1981Icar...47..470F} found that Neptune and Uranus produce Oort cloud objects much more effectively than do Jupiter or Saturn, and \citet{2013AJ....146...16L}~found this to be generally true of Neptune-mass planets vs. Jupiter-mass planets.  

To transform the odds of a single small body ever becoming a member of the Oort cloud to the fraction of the Oort cloud, we need to know the initial populations at all distances from the Sun.  The scaling usually assumed is that the surface density of the protosolar nebula followed $\Sigma \propto r^{-1.5}$, based on the mass distribution of planets today \citep{1977Ap&SS..51..153W,1981PThPS..70...35H}.  Called the Minimum Mass Solar Nebula, this construct assumes all the solid mass of the nebula ended up in planets.  To produce an initial surface density of small bodies to calculate the composition of the Oort cloud today, we assume the same scaling can be applied to the small bodies, and that all small bodies have the same size distribution.  Observations of long period comets today suggest the total mass of the Oort cloud is $1 \sim 10 m_{\oplus}$~\citep{1996ASPC..107..265W,2005ApJ...635.1348F}.  As not all Oort cloud comets have survived to the present day, this suggests the primordial mass of the solids in the Oort cloud was comparable to the total mass of solids in planets used to derive the Minimum Mass Solar Nebula.  Thus although this assumption is well-motivated\footnote{Some planet formation models expect a switch in the character of the growth when the mass of planets equals the mass of planetesimals, \citep[see][]{2004ARA&A..42..549G}}, how well it is likely to represent the correct initial conditions is not entirely clear.  Thusly caveated, we plot the total number of bodies that we infer ever make it into the Oort cloud in figure \ref{fig:oortrate} (right panel).

We simulated particles out to 50 au, but the solar nebula should have had a significant density drop-off at 30 au \citep{2004Icar..170..492G}.  Approximating this as a sharp cut-off, we find that rocky bodies from within 2.5 au of the Sun should be $\sim 4\%$~of the total population of the Oort cloud.

\begin{figure}
  \subfigure{\includegraphics[width=0.5\textwidth]{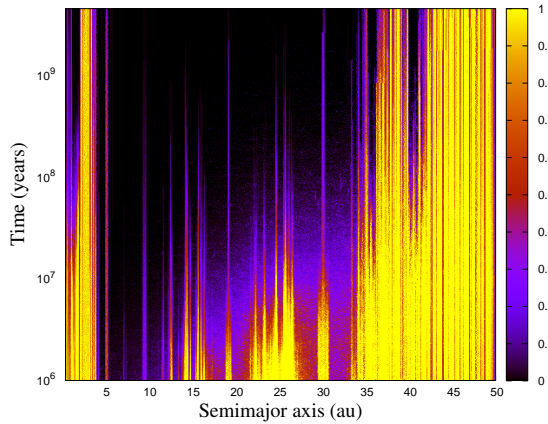}}
  \caption{Fraction of particles at a given semimajor axis, plotted against time.  The Asteroid and Kuiper belts show up strongly, as most particles are stable; the Jupiter and Neptune trojans are also visible, as well as the stable regions of the inner solar system \citep{1999Natur.399...41E}.  Two Saturnian trojans and three Uranian trojans also persist.  The concurrence between this and the observed population of small bodies is pretty good, although our non-migratory Neptune fails to capture Plutinos and other resonant objects, and the question of what eliminates the inner solar system bodies remains open.}
\label{fig:lifetimes}
\end{figure}

To better understand the fate of the particles that survive the simulations, we plot the dynamical lifetime of small bodies across the solar system in figure \ref{fig:lifetimes}.   Here we see that the asteroid and Kuiper belts are the only places in the solar system where small bodies are reliably stable for 4.5 Gyrs.  The simulations have some particles as trojans of all the planets except Mars; in the Solar system stable trojans are only known for Jupiter \citep{1907AN....174...47W}, Neptune \citep{2004MNRAS.347..833B}, and Mars \citep{2005Icar..175..397S}, but shorter lifetime trojans are also known for Venus \citep{2014MNRAS.439.2970D}, the Earth \citep{2011Natur.475..481C}, and Uranus \citep{2013Sci...341..994A}.  A couple of trojans of Uranus and Saturn surviving is not unexpected \citep{2002Icar..160..271N}, and we find 3 and 2 respectively.  Some surviving objects become scattered disk objects/centaurs.  A significant number of objects remain in the inner solar system between the terrestrial planets, a result previously found by \citet{1999Natur.399...41E,2002MNRAS.333L...1E}.  These are not observed, whatever physics caused their loss is not included in our simulations, and their absence remains an open problem.  The simulations also show a slowly decaying population between Uranus and Neptune \citep{1997Natur.387..785H}, though with only a few thousand particles, the last member is lost at $3.3 \times 10^{9}~\rm{years}$.

\section{Detectability with LSST}
\label{sec:detect}

To test this prediction, we want to find some asteroids inbound from the Oort cloud.  The number of Oort cloud comets with individual masses of $\sim 1.7 \times 10^{17}~\rm{g}$~or greater is estimated from observations to be $\sim 2 \times 10^{11}$~\citep{2005ApJ...635.1348F}.  The number of comets drops off strongly at lower masses, with only $\sim 5 \times 10^{11}$ Oort cloud comets with masses above $\sim 10^{12}~\rm{g}$~\citep{2005ApJ...635.1348F}, although the typical size and size at which the drop off occurs is somewhat uncertain \citep{1996ASPC..107..265W,2011MNRAS.414..458S,2012MNRAS.423.1674F}. Given this, when considering surveys for Oort cloud asteroids, we can approximate all comets as having the same mass, in which case the total mass of Oort cloud comets is $M_c \sim 4 \times 10^{28}~\rm{g}$, We found the total mass of Oort cloud asteroids to be $4\%$~that of comets, or $M_r \sim 2 \times 10^{27}~\rm{g}$.  Since most Oort cloud asteroids come from the terrestrial planet region, where the primordial mass distribution is uncertain, we assume a single-mass distribution for them, by default the same as comets ($m_r \sim 1.7 \times 10^{17}~\rm{g}$).  For a reasonable density for rock ($2.5~\rm{g}~\rm{cm}^{-3}$), this gives Oort cloud asteroids with radii $r_r = 2.3~\rm{km}$.  The number of Oort cloud asteroids is then $n_r = M_r/m_r \sim 8 \times 10^{9}$~rocky bodies bigger than $2.3~\rm{km}$~in the Oort cloud - {\it the Oort cloud has more asteroids than the Asteroid belt does!}

The Large Synoptic Survey Telescope (LSST) is expected to detect objects to an apparent magnitude limit of $m \sim 24.5$~\citep{2009ApJ...704..733M}.  Oort cloud asteroids with radii $2.3$~km and albedo $\alpha = 0.16$, typical of inner Main belt asteroids \citep{2014arXiv1407.4521S},~would be detectable to a distance of $b \sim 13~\rm{au}$; for other sizes and albedoes $b \propto \alpha^{0.25}r_r^{0.5}$.  Given an isotropic velocity distribution, Oort cloud comets are expected to have a linearly uniform distribution of perihelia [or $n\left(q<q_0\right) \approx q_0/a$] \citep{2004ASPC..323..371D}.  For the typical $a = 3 \times 10^4~\rm{au}$~orbit, the number of Oort cloud asteroids ever detectable is thus $n_d \sim n_r b/a \sim 3.5 \times 10^{6}$.

Over LSST's ten year mission \citep{2009ApJ...704..733M}, only a small fraction of Oort cloud asteroids will be at or near perihelion.  Assuming Oort cloud asteroids travel on orbits with $a \approx \infty$~and $e \approx 1$, they spend
\begin{equation}
 \tau = \frac{4}{3} b^{1.5} \left(2GM_\odot\right)^{-\frac{1}{2}}
\end{equation}
within a distance $b$~of the Sun.  Setting $b = 13~\rm{au}$~yields an inner solar system passage time, during which the bodies are detectable by LSST, of $\tau \sim 7$~years.  For the typical $a = 3 \times 10^4~\rm{au}$~orbit, the orbital period is $P \sim 5 \times 10^6$~years, and a typical Oort cloud asteroid spends $\tau/P \sim 10^{-6}$~of its time being detectable.  At any given time, we should expect $n_d \tau/P \sim 5$~such objects to be detectable, a number that scales as $r_r^{-7/4}$.  Additionally, with LSST expected to survey for $\sim 10$~years, new objects should enter the sphere of detection at a rate of $n_d/P \sim 0.7~\rm{yr}^{-1}$, for a total of $7$~additional objects, a number that scales as $r_r^{-5/2}$.  In total we expect $\sim 12$~such objects should be detectable, a number that drops to less than one if $r_r > 14~\rm{km}$, but will be much larger if $r_r$~is smaller than 2 km.

\section{Earth Impact Rate}
\label{sec:earthimpacts}
Earth-crossing long-period comet (LPC) nuclei pose a particular problem for impact mitigation. Whereas surveys of Near Earth Asteroids (NEAs) can enable us to identify potentially hazardous objects years or decades in advance of a possible impact, LPC nuclei are not typically discovered until they become active.  As a result most LPC nuclei are not discovered until they have crossed the orbit of Jupiter and many are not seen until they are rather closer to the Sun \citep[e.g.,][]{1994hdtc.conf..221M}. An LPC nucleus on a parabolic orbit with perihelion at less than 1 au will take less than a year to cross the distance between the orbits of Jupiter and Earth, and discovery only a few months before perihelion is common \citep{1994hdtc.conf..221M}. A nuclear detonation is probably the only plausible mitigation method that could achieve a sufficient orbital change with such a short lead time, and even then it would probably only be possible if a pre-existing delivery vehicle could be used \citep{defplaear}.

An asteroid inbound from the Oort cloud presents an even more severe warning time problem than an icy LPC nucleus, since it will (presumably) never become active and would remain as a dark rock until it appears in our skies. We might thus never know what hit us. As such it behooves us to estimate the probability of a catastrophic Oort cloud asteroid impact occurring.

LPC nuclei with perihelion $q < 1~\rm{au}$~have a $f_c \sim 2.2 \times 10^{-9}$~chance of impacting the Earth per orbit \citep{2007IAUS..236..441W}.  For the population of Oort cloud asteroids discussed in \textsection \ref{sec:detect}, this corresponds to an impact once every  $f_c P/ n\left(q < 1 \rm{au}\right)\approx 8$~Gyrs.  However, the smaller bodies we neglected there as too dim to be important for detection in surveys can be important as impactors.  For NEAs the threshold radius at which an impactor will have globally catastrophic consequences has been estimated to be 500 m. A 500 m object with a typical density of $2500~\rm{kg}~\rm{m}^{-3}$~impacting at a typical velocity of $20.9~\rm{km}~\rm{s}^{-1}$~\citep{2004Icar..170..295S} has an impact energy of $3 \times 10^{20}~\rm{J}$. By comparison typical LPC nuclei impact velocities are almost 3 times higher at around $55~\rm{km}~\rm{s}^{-1}$~\citep{1994hdtc.conf..221M,2007IAUS..236..441W}, and thus a Oort cloud asteroid of the same size (and density) will have almost an order of magnitude higher impact energy, such that the equivalent threshold size for an Oort cloud asteroid is around 250 m. Estimates of the impact frequency for NEAs larger than 500 m vary between once every 0.1 Myr \citep{2003MNRAS.346..584H} and once every 0.6 Myr \citep{2004Icar..170..295S}.

Estimates of the rate of LPC nuclei impacts are somewhat more difficult to determine. Since 1700, 6 comets per century have passed within around 0.1 au of Earth \citep[e.g.,][see also www.minorplanetcenter.net/iau/lists/ClosestComets.html]{1984AJ.....89..154S}, implying an impact rate of about once every 40 Myr \citep[a similar rate was estimated by][]{2007IAUS..236..441W}. This is
in agreement with the impact probabilities per perihelion passage calculated by \citet{1994hdtc.conf..221M} \citep[again, see][]{2007IAUS..236..441W}, for a total of 11 Earth-crossing comets per year. 11 Earth-crossing comets per year is also the frequency implied by 6 close encounters per century, which is in agreement with the bias corrections of
\citet{1967AJ.....72.1002E} and the dynamical arguments of \citet{2007IAUS..236..441W}, and indicates that the current discovery rate of around 3 per year
\citep{2012MNRAS.423.1674F} is still rather incomplete. If we use the conversion between absolute H magnitude and size of \citet{2011MNRAS.416..767S} around half of the comets used by \citet{{1984AJ.....89..154S}} have nuclei larger than 250 m in radius, and at smaller size the slopes of the size distributions of NEAs and Earth-crossing cometary nuclei are fairly similar. We thus estimate that an impact with an Oort cloud asteroid larger than 250 m in radius would likely be a once-in-a-Gyr event.  The relative contribution of Oort cloud asteroid impacts at different impact energies depends on their size distribution, and how it differs from that of comets or NEAs.  Integrated over all sizes comets account for a few percent of Earth impactors by impact energies, so we expect Oort cloud asteroids to account for $\sim 10^{-3}$~of the total impactor population.  Tunguska-like impacts, capable of smiting a city, occur every $10^3$~years \citep{2002Natur.420..294B}, and thus we expect Tunguska-sized Oort cloud asteroid impacts should occur every $\sim 10^{6}$~years.  The inferred upturn in number of impactors between $\sim 250$~and $\sim 25$~m is consistent with a collisionally evolved population \citep{1969JGR....74.2531D,2003Icar..164..334O}, which is expected for such small Oort-cloud objects \citep{2001Natur.409..589S}.

\section{Discussion}
\label{sec:discussion}

We simulate the scattering of small bodies in the Solar system, augmented with the effects of the Galactic tide and passing stars, in order to understand the fraction of Oort cloud objects that should come from the inner reaches of our planetary system.  We find that for reasonable assumptions, $\sim 4\%$~of Oort cloud objects should have come from within 2.5 au of the Sun, and hence be ice-free.  Until now, such ice-free objects have gone unnoticed, as they are much dimmer than their icy, active counterparts.  LSST, however, should be expected to find roughly a dozen of these objects.

A steeper MMSN, as is sometimes argued for \citep[e.g.][]{2007ApJ...671..878D,2013MNRAS.431.3444C} would imply the Oort cloud should contain a higher fraction of asteroids.  Similarly, particular dynamical histories of the solar system, such as the Nice model \citep{2005Natur.435..466G,2005Natur.435..459T,2005Natur.435..462M}, Grand Tack \citep{2011Natur.475..206W}, or other dynamical clearings of the asteroid belt region \citep{2002aste.conf..711P} should predict different fractions of asteroids.  Additional physics should continue to remove smaller asteroids from the asteroid belt, such as the Yarkovsky effect, and orbit changes due to collisions, and they may become members of the Oort cloud.  Different histories of the Sun's local environment \citep[e.g.,][]{1995Icar..114..258G,1997Icar..129..106F,2011Icar..215..491K} should also make different predictions.  Alternate paths for Oort cloud formation would also make different predictions \citep{2010Sci...329..187L}.  This list is not exhaustive, and this constraint would need to be combined with others to discriminate between such models.  What we present here should be considered the baseline model for comparison.

No Oort cloud asteroids have definitively been found.  However, a few possible candidates exist: \citet{1997ApJ...488L.133W} postulated an inner solar system origin for 1996 PW, but the appearance is equally compatible with it being either a D-type asteroid or an extinct comet \citep{1998Icar..132..418D,2000Icar..143..354H}.  \citet{2013A&A...555A...3K}~found that 2012 DR$_{30}$~is best fit as an A-type or V-type asteroid; both represent rocky inner-solar system material that has been differentiated.  However, due to a likely dynamical origin in the Oort cloud they dismissed the possibility that 2012 DR$_{30}$~is a rocky body formed in the inner solar system.  Other Oort cloud asteroids may be hiding among the Damacloids \citep{2005AJ....129..530J}, which have short period comet orbits but show no outgassing.  This example highlights a problem common to all the more tightly bound candidates: distinguishing an object recently escaped from the asteroid belt from an object recently arrived from the Oort cloud but perturbed onto a tighter orbit is difficult \citep[see, e.g.,][]{1997ApJ...488L.133W,2012RAA....12.1576W}.  Definitive identification is likely to lie with detecting objects during their first return to the inner solar system since their near ejection gigayears ago.  

Other sources of Oort cloud asteroids should exist - models find ejecta from the Moon-forming impact to have been of order a lunar mass~\citep{2004Icar..168..433C}; Mercury's large core and small mantle may be signatures of an impact that ejected of order a Hermian mass~\citep{2007SSRv..132..189B}; the Martian hemispheric dichotomy may be an impact basin \citep{2008Natur.453.1212A}, events that would produce a large number of small bodies. The total amount of ejecta produced by giant impacts during the formation of a terrestrial planet is estimated to be $\sim 15\%$~of the final mass \citep{2012ApJ...745...79L}, and thus we should expect to find a significant number of such bodies in the Oort cloud.  Collision ejecta should begin on higher eccentricity orbits, but the effect is not too significant; comparing our results from 0.9 au to 1.1 au to the simulations of moon-forming impact ejecta by \citet{2012MNRAS.425..657J}, re-run with eccentric planets, they find the fate of lost particles at 34\% ejection, 29\% impact the Earth, 26\% impact Venus, 9\% impact the Sun, and $<1\%$~everything else, while we find 33\% ejection, 34\% impact the Earth, 21\% impact Venus, 11\% impact the Sun, and $<1\%$~everything else.

The rarity of Oort cloud asteroids means that Earth impacts of {\ asteroids from the Oort cloud} are quite unlikely to be a threat on human timescales.  From our rate estimates, we find they should produce globally catastrophic collisions around once per Gigayear, and locally destructive impacts every million years.  Thus although evidence of such occurrences may be found in the geologic record, they are not a pressing concern for the immediate future.

\section{Acknowledgements}
We thank N.W. Evans for a useful discussion, and the referee P.R. Weissman for a review which (in particular) substantially improved the clarity of this manuscript.  AS and MW are supported by the European Union through ERC grant number 279973.  DV is supported by the European Union through ERC grant number 320964. 

\bibliographystyle{mn2e}{}
\bibliography{rockets}

\end{document}